\documentclass[english]{article}
\usepackage[no-natbib-sort]{jheppub}
\usepackage[T1]{fontenc}
\usepackage[latin9]{inputenc}
\usepackage{amsmath}
\usepackage{graphicx}
\usepackage{babel}

\newcommand{\bb}[1]{\boldsymbol{#1}}
\newcommand{\mc}[1]{\mathcal{#1}}

\newcommand{\appropto}{\mathrel{\vcenter{
  \offinterlineskip\halign{\hfil$##$\cr
    \propto\cr\noalign{\kern2pt}\sim\cr\noalign{\kern-2pt}}}}}

\title{Pavlov Learning Machines}
\author[a,b]{Elena Agliari,}
\author[a,b]{Miriam Aquaro,}
\author[b,c]{Adriano Barra}
\author[a,b]{Alberto Fachechi}
\author[a,b]{Chiara Marullo}
\affiliation[a]{Dipartimento di Matematica ``Guido Castelnuovo'', Sapienza Universit\`a di Roma, Roma, Italy}
\affiliation[b]{GNFM-INdAM, Gruppo Nazionale di Fisica Matematica, Istituto Nazionale di Alta Matematica, Italy}
\affiliation[c]{Dipartimento di Matematica e Fisica ``Ennio De Giorgi'', Universit\`a del Salento, Lecce, Italy}
\emailAdd{adriano.barra@unisalento.it}

\abstract{As well known, Hebb's learning traces its origin in Pavlov's Classical Conditioning, however, while the former has been extensively modelled in the past decades (e.g., by Hopfield model and countless variations on theme), as for the latter modelling has remained largely unaddressed so far; further, a bridge between these two pillars is totally lacking. The main difficulty towards this goal lays in the intrinsically different scales of the information involved: Pavlov's theory is about correlations among \emph{concepts} that are (dynamically) stored in the synaptic matrix as exemplified by the celebrated experiment starring a dog and a ring bell; conversely, Hebb's theory is about correlations among pairs of adjacent neurons as summarized by the famous statement {\em neurons that fire together wire together}. 
In this paper we rely on stochastic-process theory and model neural and synaptic dynamics via Langevin equations, to prove that -- as long as we keep neurons' and synapses' timescales largely split -- Pavlov mechanism spontaneously takes place and ultimately gives rise to synaptic weights that recover the Hebbian kernel. 
}

\begin{document}
\maketitle

\section{Introduction}

The ``Cell Assembly Theory'', introduced in 1949 by Donald Hebb in his milestone {\em The Organization of Behaviour} \cite{Hebb}, looks at learning processes as collective features of neurons and can be  effectively summarized by the famous prescription ``neurons that fire together wire together''. Since then, Hebb's learning has become a  prolific cornucopia for modeling neural networks \cite{CKS,Nishimori}. In particular, the Hopfield model \cite{AGS-PRL}, where the Hebbian scheme is implemented mathematically, constitutes a paradigmatic model which has been attracting much attention since the mid-eighties, following the seminal investigation by Amit, Gutfreund and Sompolinsky \cite{AGS-PRL} based on statistical-mechanics tools. Indeed, after the breakthroughs by Parisi \cite{Giorgio} and coworkers  \cite{COV,MPV} at the turn of the seventies in the eighties, statistical mechanics of disordered systems was established as the theoretical reference where spontaneous information processing capabilities of Hebbian-like neural networks can be fruitful inspected.
Along this way important achievements have been collected (see e.g.~ \cite{amit1989,CKS}), and nowadays Hebb's rule is usually the starting point when getting acquainted with (bio-inspired) neural networks and related modelling. However, Hebb's rule did not appear out of the blue, rather it was an enlightened synthesis of decades of meticulous inspections in neurophysiology and psychology, whose origins have to be traced back to the
systematic studies of  Ivan Pavlov on {\em Classical Conditioning}\footnote{It is worth noting that, roughly in the same years, also Thorndike \cite{Twenty} and Twitmyer \cite{PavlovUSA} understood the core idea behind Classical Conditioning  but they drew minor attention from the scientific community at that time.}, which greatly inspired Donald Hebb\footnote{It is reported that Boris Babkin, a student of Pavlov's, spurred Donald Hebb to deepen Classical Conditioning.}. 
Despite this closeness, the two theories -- Hebb's learning based on neuron-activity correlation and Pavlov's learning based on conditioning by stimuli -- had quite distant fates, at least in the context of theoretical modelling, the former being by far much more represented in the mathematical and physical literature (as for the latter, one should still recall significant contributions by Estes \cite{cinquantino}, Hull and Spence \cite{ReviewP}, and modern extensions by Rescorla-Wagner \cite{ResWag} and Mackintosh \cite{Macka}). Hereafter we aim to provide a mathematical description of Classical Conditioning, whence one could appreciate how {\em Pavlov implies Hebb}: this bridge is the main focus of the present work\footnote{Inspiring ideas about this connection between Pavlov and Hebb from a modeling perspective are due to Francesco Guerra, see \cite{GuerraRoberto,GuerraEBV}.}.

Remarkably, this requires a smooth interpolation from correlations at the neural level up to correlations among concepts, thus connecting neurophysiological to behavioral perspectives. In fact, as anticipated, Hebb's learning rule suggests that if two adjacent neurons are emitting spikes, in a temporally correlated manner, the {\em wire} connecting them should be enlarged\footnote{And, in some other parts of the network, other under-used wires should be diminished in order to preserve network's omeostasis, ultimately approaching an optimal current flow criterion.}; on the other hand, Pavlov's learning rule connects {\em concepts}, somehow preserving the same meaning, namely, if the two stimuli are simultaneously presented for a long enough time (\emph{e.g.}, for a timescale larger than the characteristic synaptic timescale), then the synapses rearrange in order to store this correlation (as food presentation and bell's ringing in the famous example). Thus, in our framework, we ultimately need to interpolate between different degrees of information, from the simple bit managed by the single neuron to patterns of bits codifying for concepts handled by neural circuits.

In general, this procedure is not unique, rather it can be achieved in countless ways, and here we will present a simple scheme which has the advantage to be adaptable to different contexts. Indeed, we will think of the neurons as the nodes of a network whose weighted links represent the synapses: the fact that solely by  assuming that neural and synaptic timescales are quite spread apart suffices to obtain Classical Conditioning as a spontaneous phenomenon makes this model a rather general one not confined to neural assemblies. This observation suggests that large, frustrated\footnote{With the term {\em frustration} we mean that the nodes in the network can be coesive but also competitive as the links connecting them can assume both positive as well as negative values, as for instance excitatory vs inhibtory synapses in neural networks, eliciting or suppressive cytokines in lymphocyte networks  \cite{CoolenSaturation,Germain,GiorgioImmune}, etc.} networks of interacting units can spontaneously exhibit computation capabilities, as typically happens broadly in the biological world, from the inter-cellular level (\emph{e.g.} lymphocyte networks within the immune system \cite{Immune}) to the intra-cellular level (\emph{e.g.}  gene regulatory networks \cite{Gene}), for invertebrates \cite{inverter} as well as vertebrates \cite{diretto} and even  for prokaryotes as well as  eucaryotes \cite{PNASSO}. 

The paper is structured as follows. In Sec.~\ref{sec:SM} we present the theoretical framework to embed our model and we derive a set of stochastic evolution equations for the neuronal configuration and the synaptic arrangement. Then, in Sec.~\ref{sec:IV}, we study the related dynamics recovering the Classical and Generalized Conditioning for ``simple'' concepts and show that, in the long-time limit, the synaptic matrix displays a Hebb-like shape. Finally, Sec.~\ref{sec:V} is left for discussions and outlooks, while technical details and further insights are collected in the Appendices.


\section{Neural and synaptic dynamics derived from statistical mechanics} \label{sec:SM}
\subsection{A teaspoon of Statistical Mechanics for neural networks}

We consider a network made up of $N$ binary neurons $\sigma_i=\pm 1$ for each $i=1,\dots,N$. The configuration space is denoted as $\Sigma_N=\{-1,+1\}^N$ and each configuration of the system is represented by the corresponding point $\bb\sigma = \{\sigma_1, ..., \sigma_N\} \in \Sigma_N$. We assume that the network cost function is described by the 2-point Hamilton function
\begin{equation}
{H}_{N,\bb J}(\bb \sigma)=-\frac{1}{2}\sum_{i\neq j=1}^{N,N}J_{ij}\sigma_{i}\sigma_{j}-u\sum_{i=1}^N h_{i}\sigma_{i},
\end{equation}
where $\bb J =(J_{ij})$ is the (symmetric) coupling (referred to as the {\em synaptic matrix}) describing pairwise interactions between the neurons, $h_i\in\{-1,0,+1\}$ is the bias (or {\em firing threshold} in more biological terms) acting on each neuron and $u$ is a global amplification factor (the strength of the bias). Once the Hamilton function is introduced, we can equipe the configuration space $\Sigma_N$ with a Boltzmann-Gibbs measure whose associated partition function is
$$
Z_{N,\beta, \bb J}=\sum_{\bb\sigma\in \Sigma_N} \exp\big[-\beta H_{N,\bb J}(\bb\sigma)\big],
$$
where we introduced $\beta$ to account for the noise, such that $\beta \to \infty$ is the noiseless limit while $\beta \to 0$ corresponds to a fully random system.
For fixed coupling matrix $\bb J$, the probability distribution governing the equilibrium dynamics in the configuration space is thus given by
$$
\mc P_{N, \beta, \bb J}(\bb \sigma)=\frac1{Z_{N,\beta, \bb J}} \exp\big[-\beta H_{N,\bb J}(\bb\sigma)\big].
$$
\par\medskip
In disordered systems, the presence of random interactions among the constituting units (see \emph{e.g.}, \cite{CKS}) implies that the coupling matrix is a random variable, with the crucial problem being the computation of the {\it quenched} free-energy, {\it i.e.} the expectation of the free-energy $F_{N, \beta, \bb J}:=-\frac{1}{\beta} \log Z_{N,\beta, \bb J}$ w.r.t. the probability distribution of the weights $\bb J$. This is motivated by the existence of self-averaging theorems \cite{ShcherbinaPastur-JSP1991,Bovier-JPA1994} ensuring that the fluctuations of the free-energy w.r.t. its expectation value vanish in the thermodynamic limit $N\to\infty$. 
In a nutshell, the quenched free-energy paints a physical situation where we study the thermalization of the neurons at fixed (but random) weights $J_{ij}$, thus averaging over all their possible realizations {\it after} taking the logarithm of the partition function: this procedure intrinsically matches the so-called {\it adiabatic hypothesis}, prescribing that the coupling evolution, if any, takes place on much longer time-scales w.r.t. the neural dynamics, such that couplings can be considered as fixed (or {\em quenched}) when addressing the neural dynamics.
\par\medskip
Let us now consider a situation where both degrees of freedom, $\bb \sigma$ and $\bb J$, evolve but fthe related time-scales are different, say $\bb J$ is much slower than $\bb \sigma$ (as it is the case for synapses and neurons, see \emph{e.g.}, \cite{Tuckwell-1988}), but the experiment is run for a time long enough to appreciate the evolution of both. Then, we need to take into account both neural and coupling dynamics, that is, we need to refer to the {\it joint} probability distribution $\mc P_{N,\beta}(\bb\sigma,\bb J)=P_{N,\beta}(\bb\sigma\vert \bb J)P(\bb J)$, with $P(\bb J)$ being the prior distribution for the coupling matrix, and trivially $P_{N,\beta}(\bb\sigma\vert\bb J)\equiv P_{N,\beta,\bb J}(\bb\sigma)$. Thus, the central quantity we deal with is
\begin{equation}
	\label{eq:Z_0}
	Z_{N,\beta}= \sum_{\bb\sigma\in\Sigma_N} \int dP(\bb J)\exp\big(-\beta H_{N,\bb J}(\bb\sigma)\big),
\end{equation}
where $ dP(\bb J)$ is the measure w.r.t. the prior distribution of the weights. In statistical mechanics, this is formally related to the {\it annealed} free energy. In the following we retain this setup and we  obtain a system of differential equations governing the dynamics of the whole model \cite{Gerstner,Murray}.

In the next Sec.~\ref{ssec:rad}, we  consider the case of a coupling matrix whose entries are drawn  i.i.d. from a Rademacher prior and whose evolution follows a Langevin-like dynamics to show that the long-term relaxation for synapses matches the mathematical formulation of Hebb's learning rule, further, we show numerically that Classical Conditioning is also recovered. These results shall be generalized and extended in the following Sec.~\ref{sec:IV}.

\subsection{Neural and synaptic Langevin dynamics  from Rademacher priors} \label{ssec:rad}
In this case, the space of interaction matrices is endowed with the probability measure
\begin{equation}
P(J_{ij}=\pm1)=1/2, ~ \textrm{for any}~ i,j =1, ..., N, ~ \textrm{and}~ i\neq j.
\end{equation}
Before proceeding, it is computationally convenient to recast the partition function \eqref{eq:Z_0} with the addition of source terms $\bb t$ and $\bb T$ for, respectively, the neurons and the weights as
\begin{equation}
{Z}_{N, \beta}(\bb t, \bb T)=\sum_{\bb \sigma}\sum_{\bb J}\exp\Big(\beta\sum_{i<j}J_{ij}\sigma_{i}\sigma_{j}+\beta u\sum_{i}h_{i}\sigma_{i}+\sum_{i}t_{i}\sigma_{i}+\sum_{i<j}T_{ij}J_{ij}\Big),
\end{equation}
with
$$
\sum_{\bb J} \equiv \prod _{i<j} \sum_{J_{ij}=\pm1}.
$$
In fact, by performing the sum over all $J_{ij}$ variables, we get
\begin{equation}
{Z}_{N,\beta}(\bb t, \bb T)=\sum_{\bb\sigma}\exp\Big(\sum_{i}\left(\beta  h_{i}u+t_{i}\right)\sigma_{i}\Big)\prod_{i<j}2\cosh\left(\beta\sigma_{i}\sigma_{j}+T_{ij}\right),
\end{equation}
and, by taking the derivative w.r.t. the source $T_{ij}$, we have
\begin{equation} \label{eq:a}
\langle J_{ij}\rangle=\frac{\partial\log {{Z}_{N,\beta}( \bb t, \bb T)}}{\partial T_{ij}}\Big\vert_{\bb T=0, \bb t =0}=\langle\tanh\left(\beta\sigma_{i}\sigma_{j}\right)\rangle=\langle\sigma_{i}\sigma_{j}\rangle\tanh\beta \underset{m.f.}{\approx} \langle\sigma_{i} \rangle \langle \sigma_{j} \rangle  \tanh\beta ,
\end{equation}
where we used the fact that $\tanh (\sigma x ) = \sigma \tanh x$ for $\sigma=\pm1$ and, in the last passage, 
{\it m.f.} stands for mean-field assumption (which is expected to hold $\beta$-almost everywhere in the thermodynamic limit): this approximation consists in the fact that each neuron is subject to a net effective external field (rather than the mutual interaction with other neurons), so that relevant correlation functions factorize in that limit. 
\newline
Similarly, the expectation value of the neural activity $\sigma_i$ is obtained as 
\begin{equation} \label{eq:b}
\langle\sigma_{i}\rangle=\frac{\partial\log {{Z}_{N, \beta}(\bb t, \bb T)}}{\partial t_i}\Big\vert_{\bb T=0, \bb t =0}=\langle\tanh\big(\beta\sum_{j\neq1}J_{ij}\sigma_{j}+\beta h_{i}u\big)\rangle\underset{m.f.}{\approx}\tanh\big(\beta\sum_{j\neq i}\langle J_{ij}\rangle\langle\sigma_{j}\rangle+\beta h_{i}u\big),
\end{equation}
where we used again the mean-field assumption. 
\newline
Since all neural indexing is equivalent, the equalities \eqref{eq:a}-\eqref{eq:b} hold \emph{mutatis mutandis} for each $i=1,\dots,N$ and $j=1,\dots,N$ with $i \neq j$. 
\newline
With these results, we can set up a Langevin-like dynamics governed by the following evolutive equations:
%
\begin{eqnarray}
	\dot{\langle{\sigma_{i}}\rangle}&=&-\frac{1}{\tau}\langle\sigma_{i}\rangle+\frac{1}{\tau}\tanh\big(\beta\sum_{j\neq i}^{N}\langle J_{ij}\rangle\langle\sigma_{j}\rangle+\beta h_{i}u\big),\label{eq:sigma_rad}\\
	\dot{\langle{J_{ij}}\rangle}&=&-\frac{1}{\tau^{\prime}}\langle J_{ij}\rangle+\frac{1}{\tau^{\prime}}\langle\sigma_{i}\rangle\langle\sigma_{j}\rangle\tanh\beta,\label{eq:J_rad}
\end{eqnarray}
where we introduced $\tau$ and $\tau'$ as, respectively, the neural and synaptic time-scales with $\tau \ll \tau'$. 
\newline
It is worth pointing out that the long-term relaxation of eq. (\ref{eq:J_rad}), that is obtained by imposing  $\dot{\langle{J_{ij}}\rangle}=0$, yields $\langle J_{ij} \rangle = \tanh(\beta)\langle \sigma_i \rangle \langle \sigma_j \rangle$ such that, by applying two single-bit stimuli $\xi_i$ and $\xi_j$ on the two neurons $\sigma_i$ and $\sigma_j$ (\emph{i.e.}, constraining  $\sigma_i = \xi_i$ and $\sigma_j = \xi_j$), we recover the Hebbian structure for the synaptic matrix, namely
\begin{equation}\label{eq:JJ}
J_{ij} = \tanh (\beta) \xi_i \xi_j,
\end{equation}
where the noise term $\beta$ shifts the coupling intensity: in the $\beta \to 0$ limit the noise prevails over the stimuli and no learning can be accomplished, vice versa in the opposite limit. 
Remarkably, the boundedness of $J_{ij}$, that emerges from our model ($J_{ij}\in [-1,+1]$, from eq.~\eqref{eq:JJ}), is consistent with the early investigations on Classical Conditioning (see \emph{e.g.}, the Rescorla-Wagner \cite{ResWag} and Mackintosh \cite{Macka} models). In particular, it was noticed that a mandatory requisite for a successful conditioning is the usage of bounded synapses such that, once they reach their upper bounds, no effect is produced any longer, that is, {\em blocking} takes place. 
Nonetheless, it is instructive to work out the same calculations by using non-bounded variables for the synapses: as shown in Appendix \ref{ssec:gauss}, by assuming Gaussian priors for the synapses, we end up with the same learning rule but $\tanh (\beta)$ is replaced with $\beta \in (0, +\infty)$.
%

\bigskip

To see this theoretical picture at work we perform extensive simulations and, for the sake of transparency, hereafter we report the discretized version of the Langevin dynamics \eqref{eq:sigma_rad}-\eqref{eq:J_rad} for numerical implementation:
\begin{eqnarray} \label{discrete_time_a}
	\sigma_{i}^{(n+1)}&=&\sigma_{i}^{(n)}\left(1-\frac{\delta t}{\tau}\right)+\frac{\delta t}{\tau}\tanh\big(\beta\sum_{i\neq j}^{N}J_{ij}^{(n)}\sigma_{j}^{(n)}+u\beta h_{i}^{(n)}\big),\\
	 \label{discrete_time_b}
	J_{ij}^{(n+1)}&=&J_{ij}^{(n)}\left(1-\frac{\delta t}{\tau^{\prime}}\right)+\frac{\delta t}{\tau^{\prime}}\sigma_{j}^{(n)}\sigma_{i}^{(n)}\tanh\beta,
\end{eqnarray}
where we posed $\langle J_{ij}\rangle=J_{ij}$
and $\langle\sigma_{i}\rangle=\sigma_{i}$ to lighten the notation, $\delta t$ identifies the time unit for the system relaxation, and $n$ represents the number of time steps elapsed (iterations), such that $t = n \delta t$; we also stressed the dependence of the fields $h_i^{(n)}$ on the time step $n$, since in general the external stimulus is presented for a limited temporal window. 
 %
 
 \begin{figure}[tb]
\begin{centering}
\includegraphics[width=0.75\textwidth]{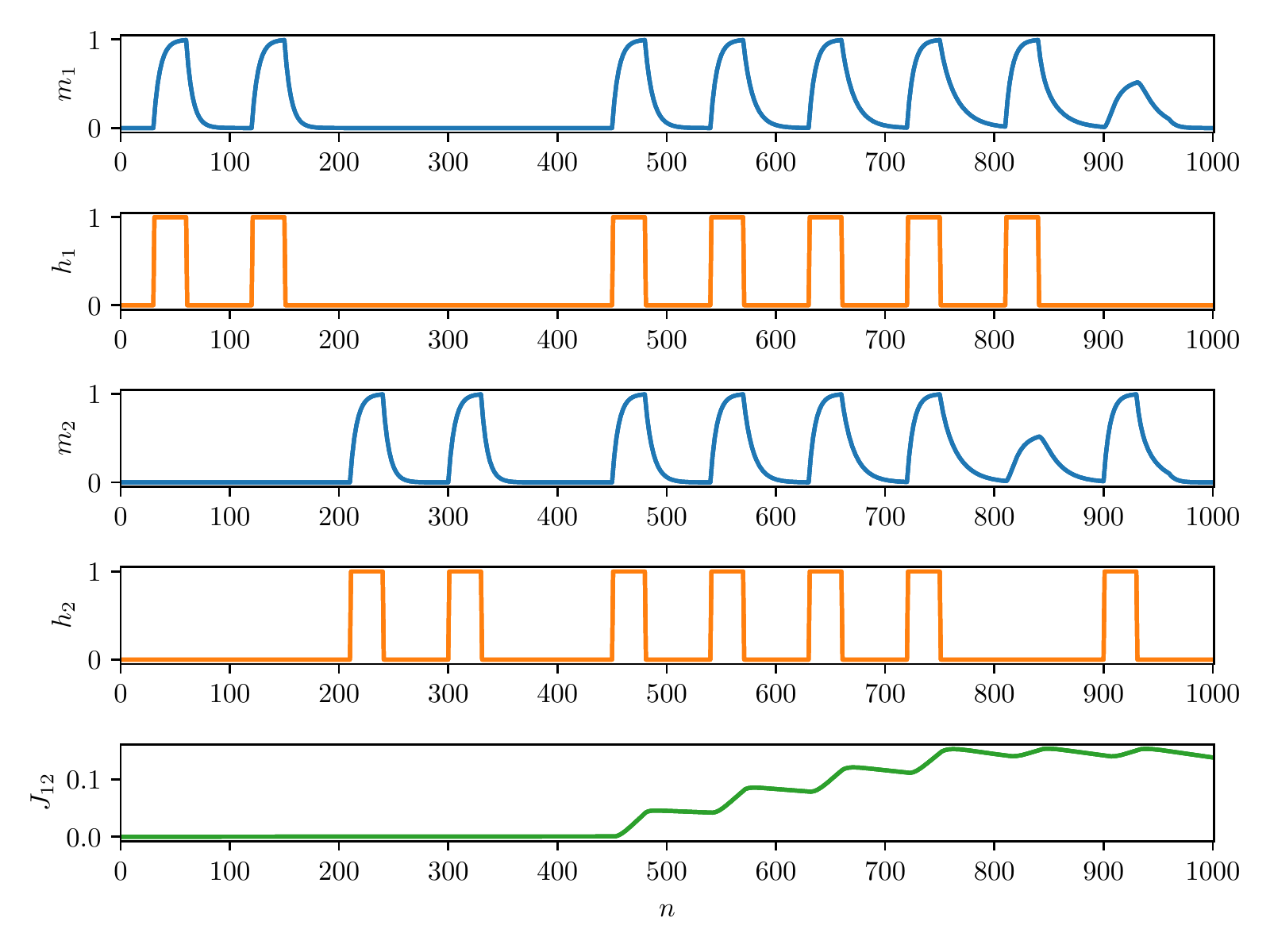}
\par\end{centering}
\caption{\label{figtest}{\bfseries Classical conditioning via Langevin dynamics on single couples of neurons}. The five panels report the evolution of the neural activities through $m_{i}^{(n)}\, {\rm with}\,\, i=1,2$ (blue lines), of  the stimuli presented to the network through $\bb h^{(n)}\cdot \bb h^{i}\,{\rm with}\,\, i=1,2$ (orange lines, which we indicated with $h_{1,2}$ to lighten the notation), and of the synaptic matrix $J_{12}$ (green line).  During the first $450$ iterations the stimuli are presented separately and we check that the network responds properly. In the next $300$ iterations the stimuli are simultaneously presented to the network and, correspondingly, their correlation starts to get stored in the relative coupling (the green curve raises from zero). As a result, when, at $n=800$, we present solely the first stimulus, both the magnetizations $m_1$ and $m_2$ respond, confirming that classical conditioning has successfully taken place. We used as parameters of the simulation to be inserted in equations \eqref{discrete_time_a},\eqref{discrete_time_b}: $\beta=4$, $N=2$, $\delta t=0.16\tau$,
$\tau/\tau^{\prime}=0.012$ and $u=1$. In this experiment the time length of any stimulation is $n_{s}=30$ iterations in order for the neurons to have sufficient time to align themselves with it, usually this happens when the stimulus is presented for a time $t\geq\tau$ which in terms of number of iterations corresponds to $n_{s}\delta t \geq\tau$, and since in the simulation $\delta t=0.16 \tau$, $n_{s}$ must be $\gtrsim 10$. }
\end{figure}

In particular, in our first numerical experiment, we take $N=2$ and the stimuli presented to the neurons are $\xi^{1}\in\{-1,1\}$ for neuron $1$ and $\xi^{2}\in\{-1,1\}$ for neuron $2$. The field $\bb h^{(n)}$ inserted in eqs.~\eqref{discrete_time_a}-\eqref{discrete_time_a} takes the form 
\begin{equation}\label{guerrastyle}
\boldsymbol h^{(n)} = 
\begin{cases}
\bb h^{1}:=(\xi^{1},0) ~~ \textrm{for} ~ n \in \mathcal I_1\\
\bb h^{2}:=(0,\xi^{2}) ~~ \textrm{for} ~ n \in \mathcal I_2\\
\bb h^{3}:=(\xi^{1},\xi^{2}) ~~ \!\textrm{for} ~ n \in \mathcal I_3\\
(0,0)~~~~ \textrm{elsewhere}
\end{cases},
\end{equation}
where $\mathcal I_i$ with $i=1,2$ are the intervals of time steps where the neuron $i$ occurs to be stimulated whereas $\mathcal I_3$ is the interval of time steps where both neurons occur to be stimulated.
To check the alignment between the neural configuration $\boldsymbol \sigma^{(n)}$ and the stimulus $\boldsymbol h^{(n)}$ we also introduce the {\em magnetization} $\bb m^{(n)}$:
\begin{equation}\label{magsti}
m_{i}^{(n)}=\frac{\boldsymbol\sigma^{(n)} \cdot \bb h^{i}}{\left\Vert \bb h^{i}\right\Vert _{1}}=\sigma_{i}\xi^{i}, \,\,{\rm where}\,\,\left\Vert \bb h^{i}\right\Vert _{1}=\sum_{j=1}^{2}|h^{i}_{j}|, \quad i\in\{1,2\}
\end{equation}
that measures the alignment of the neuron $i$ with the related stimulus $\xi^{i}$.
Results for the time evolution of $\bb h^{(n)},  \bb m^{(n)}$, and $J_{12}^{(n)}$ are shown in Fig.~\ref{figtest}. In particular, in the first part of the experiment the two neurons are stimulated separately and each single stimulation session corresponds to an alignement of the related neuron, while the other neuron does not exhibit any persistent orientation giving rise to a null magnetization, the synapse $J_{12}$ also remains neutral. In the second part of the experiment the two neurons are stimulated simultaneously, and they both react accordingly: this time the synapse $J_{12}$ is also affected and stores their correlation. This information is retained by the system even when both fields are switched off in such a way that, when, in the third part of the experiment only one neuron is stimulated, also the second one reacts.  
The system has therefore learnt to relate the stimuli: solely one of the two is sufficient to prompt the retrieval of both $\xi^{1}$ and $\xi^{2}$.

\section{From Pavlov's Classical Conditioning to Hebb's learning} \label{sec:IV}

Once understood that a multi-scale Langevin dynamics can naturally lead to Pavlov's Conditioning at the single synapse level, we enlarge the setting moving from stimuli made of simple bits to stimuli made of concepts represented by {\em patterns} and therefore employing a large number of neurons and synapses. The simplest path consists in considering two stimuli affecting different neural areas as in Classical Conditioning theory and this will be the first scenario adressed hereafter in Sec. \ref{sec:pavlov}; later in Sec.~\ref{sec:GC} we will generalize this setting by considering several stimuli at once toward modern versions of Ceneralized Conditioning as those studied by Rescorla-Wagner \cite{ResWag} or by Mackintosh \cite{Macka}. Next, in Sec.~\ref{sec:Hebb} we will show the long-time limit of the coupling matrix subjected to different stimuli displays a Hebbian structure and finally in Sec.~\ref{sec:obs} we also discuss phenomena like obsessions and unlearning.  

\subsection{Emergence of concept's correlations by Classical Conditioning}\label{sec:pavlov}
In this subsection the plan is, first, to make the network learn two uncorrelated patterns by presenting them separately and checking the absence of correlations, namely, when we re-present one of them, after learning, solely that pattern is retrieved, as expected in Hebb's learning; then, we present both patterns simultaneously and persistently (\emph{i.e.}, for a time window larger than the synaptic timescale) and we inspect how these patterns get correlated within the synaptic matrix, such that, after learning, by presenting solely one of them, the network actually retrieves both of them, confirming a successful  Conditioning.
\newline
To start with this plan on our network built of by $N$ binary neurons, we introduce two stimuli $\boldsymbol \xi^1$ and $\boldsymbol \xi^2$ as two vectors of length $N$ to be applied to the system. As mentioned above, here we choose a particular structure for stimuli which follows from the original Pavlov setting, where stimuli come by different sensing involving different neural regions in the brain (\emph{e.g.}, visual for the food and acoustic for the ringing bell). Thus, we assume that the two stimuli involve different neurons and, in the simplest setting, both the concepts to learn stimulate exactly $N/2$ neurons\footnote{The general case of
patterns of arbitrary length is deepened in Appendix B.}. Specifically, $\boldsymbol \xi^1$ and $\boldsymbol \xi^2$ are arranged as
\begin{eqnarray}
\boldsymbol \xi^1&=& (\xi_1^1, \xi_2^1, ..., \xi^1_{\frac{N}{2}},0,0,...,0)\\
\boldsymbol \xi^2&=& (0,0,...,0, \xi^2_{\frac{N}{2}+1}, \xi^2_{\frac{N}{2}+2}, ..., \xi^2_N).
\end{eqnarray}
%
The fields inserted in \eqref{discrete_time_a}-\eqref{discrete_time_b} and stemming from these stimuli read as:
\begin{equation}\label{fieldvar}
h^{(n)}_{i} = 
\begin{cases}
\xi^{(n)}_{i}&\textrm{if}~~|\xi^{(n)}_{i}|=1\\
U(\{-1,+1\})&\textrm{otherwise}
\end{cases}
\end{equation}
with
\begin{equation}
\boldsymbol{\xi}^{(n)} = 
\begin{cases}
\boldsymbol \xi^1 ~~ \textrm{for}   ~~ n \in \mathcal I_1\\
\bb \xi^2 ~~ \textrm{for}  ~~ n \in \mathcal I_2\\
\end{cases}.
\end{equation}
where $\mathcal I_{\mu}$ with $\mu=1,2$ represents the set of time sectors where the stimulus containing $\boldsymbol \xi^{\mu}$ is active and $U(\{-1,+1\})$ is a Rademacher random variable.
As anticipated, in the early learning stage, we present to the network these patterns separately until these are learnt and, at the end of this trivial exercise, the resulting synaptic matrix of the network convergences to  a modular Hebbian network  \cite{Hierarchical,ScaleFree}, whose blocks share the same size $\frac{N}{2}\times\frac{N}{2}$ and the off-diagonal blocks are null:
\begin{equation}\label{scorrelata}
\bb J(\{\boldsymbol \xi^1, \boldsymbol \xi^2 \})=\frac{1}{2}\times\left(\begin{array}{cc}
\boldsymbol{J}^{1} & \boldsymbol{0}\\
\boldsymbol{0} & \boldsymbol{J}^{2}
\end{array}\right),
\end{equation}
namely
\begin{equation}\label{factor12}
J_{ij}(\{\boldsymbol \xi^1, \boldsymbol \xi^2 \})=\frac{1}{2}\times\begin{cases}
\xi^1_{i}\xi^1_{j} & (i,j)\in B_{1}\\
\xi^2_{i} \xi^2_{j} & (i,j)\in B_{2}\\
~0 & (i,j)\vee (j,i)\in B_{mix}\\
\end{cases}.
\end{equation}
where
$
B_{1}=\left[1,N/2\right]\times\left[1,N/2\right],\, B_{2}=\left[N/2+1,N\right]\times\left[N/2+1,N\right],\, B_{mix}=\left[1,N/2\right]\times\left[N/2+1,N\right].
$
In order to understand why we expect the synaptic matrix to converge precisely to the structure coded in equation \eqref{factor12}, we use the result obtained in Appendix \ref{sec:fluct}: in the end of the training, the synaptic matrix converges to the temporal mean of the stimuli $J_{ij}(t)=\frac{1}{t}\int_{0}^{t}h_{i}(\tau)h_{j}(\tau)d\tau
$ plus a fluctuation which depends on the noise and on the ratio $\tau/\tau^{\prime}$ (see equation \eqref{hebb_dim}). In this case there are two patterns given as stimuli with equal frequency during the stimulation sessions,\emph{i.e.}\,$\,\bb h^{(n=1)},\bb h^{(n=2)}$, thus the temporal mean of the stimuli is equal to $\frac{1}{2}\sum_{k=1}^{2}\xi_{i}^{k}\xi_{j}^{k}$ which corresponds to \eqref{factor12}.\\
As for the off-diagonal blocks, since there are no correlations among the related neurons, they contain no information.

Next, we present both the stimuli simultaneously, namely the stimulus presented to the network is 
\begin{equation}
\boldsymbol \xi^{1+2}=  \boldsymbol \xi^1 + \boldsymbol \xi^2 = (\xi_1^1, \xi_2^1, ..., \xi_{\frac{N}{2}}^1,\xi^2_{\frac{N}{2}+1}, \xi^2_{\frac{N}{2}+2}, ..., \xi^2_N).
\end{equation}
and it is retained for long enough to allow synapses to be plastic. The resulting Hebbian matrix converges to  
\begin{equation}
\bb J(\bb \xi^{1 + 2})=\left(\begin{array}{cc}
\boldsymbol{J}^{1} & \boldsymbol{J}^{\,\rm mix}\\
\boldsymbol{J}^{\,\rm mix} & \boldsymbol{J}^{2}
\end{array}\right),
\end{equation}
namely\footnote{Let us note that the $1/2$ factor is no longer present: this is because only one pattern is given as stimulus (in the form $\bb\xi^{1+2}$) and the network learns only this one.}
\begin{equation} \label{eq:J_H}
J_{ij}(\boldsymbol \xi^{1+2})=\begin{cases}
\xi^1_{i} \xi^1_{j} & (i,j)\in B_{1}\\
\xi^2_{i} \xi^2_{j} & (i,j)\in B_{2}\\
\xi_{i}^1 \xi_{j}^2 & (i,j)\in B_{mix}\\
\xi^1_{j} \xi^2_{i} & (j,i)\in B_{mix}
\end{cases}.
\end{equation}
\par
\begin{figure}
\begin{centering}
\includegraphics[scale=0.7]{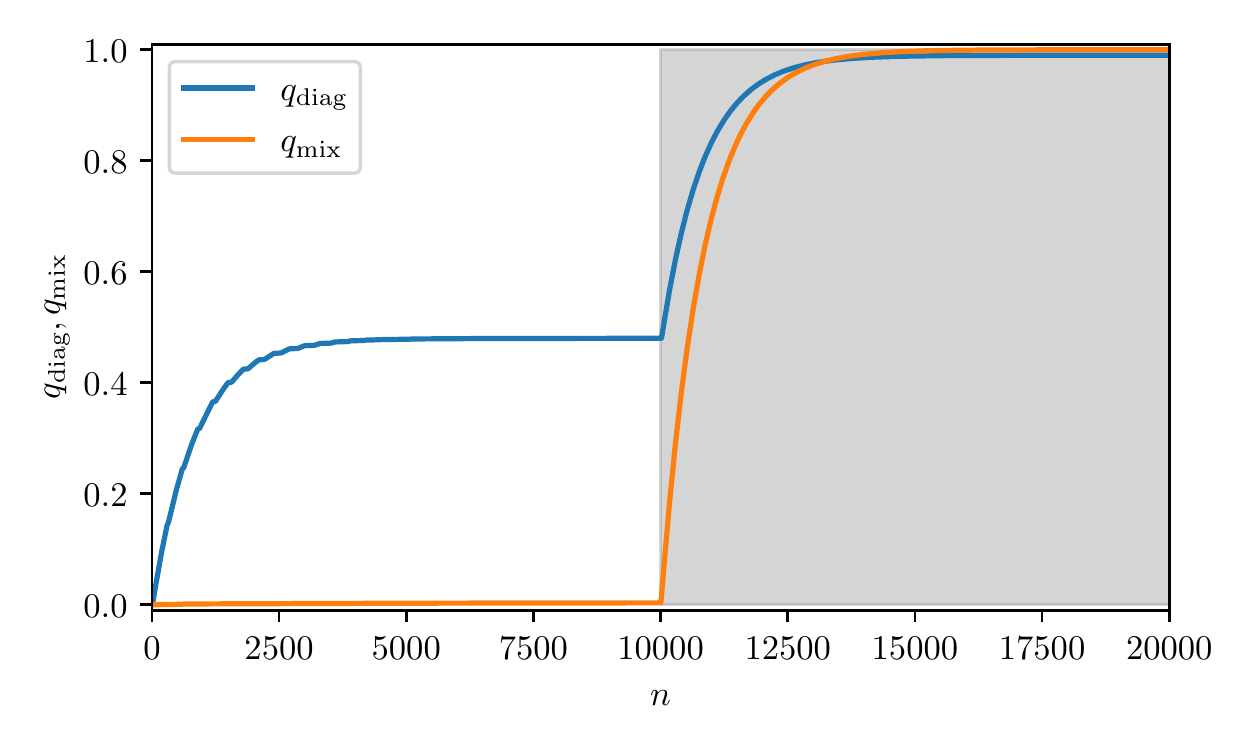}
\par\end{centering}
\caption{\label{fig1}\textbf{Evolutions of the overlaps.} The overlaps $q_{\rm diag},\,q_{\rm mix}$  defined in eq. \ref{q_planted} versus the time step $n$: in the left side of the plot (white background) the two stimuli are presented separately and alternately to the network, in particular every sigle stimulus is presented consecutively for $n=300$ iterations,whereas in the right side of the plot (grey background) the two stimuli are jointly presented. Note that in the left side of the plot (for $n\leq 10000$) standard learning of patterns takes place, instead, in the right side of the plot (i.e. for $n> 10000$) classical conditioning takes place too. We used as parameters of the simulation to be inserted in equations \eqref{discrete_time_a},\eqref{discrete_time_b}: $N=200$, $\beta=10$, $\tau/\tau^{\prime} = 0.012$, $\delta t=0.10\tau$ and the strength of the stimuli is $u=200$.}
\end{figure}

\begin{figure}
\begin{centering}
\includegraphics[width=0.95\columnwidth]{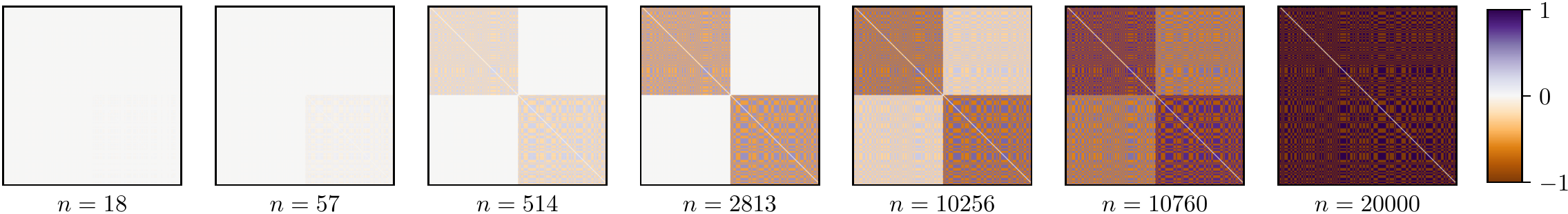}
\par\end{centering}
\caption{\label{Jmix}\textbf{Evolution of the synaptic matrix when only two stimuli are involved.\,}
Evolution of the synaptic matrix for several iterations: for $n\leq 10000$  the synaptic matrix evolves while storing the single patterns (pure learning mode of single stimuli)  while the last three images (for $n> 10000$) show how classical conditioning forces the off-diagonal blocks to have non-null entries accounting for the Pavlovian correlation among concepts. The parameters of the simulation are the same as Fig.\,\ref{fig1}.}
\end{figure}

\begin{figure}[tb]
	\centering
	\includegraphics[width=0.9\textwidth]{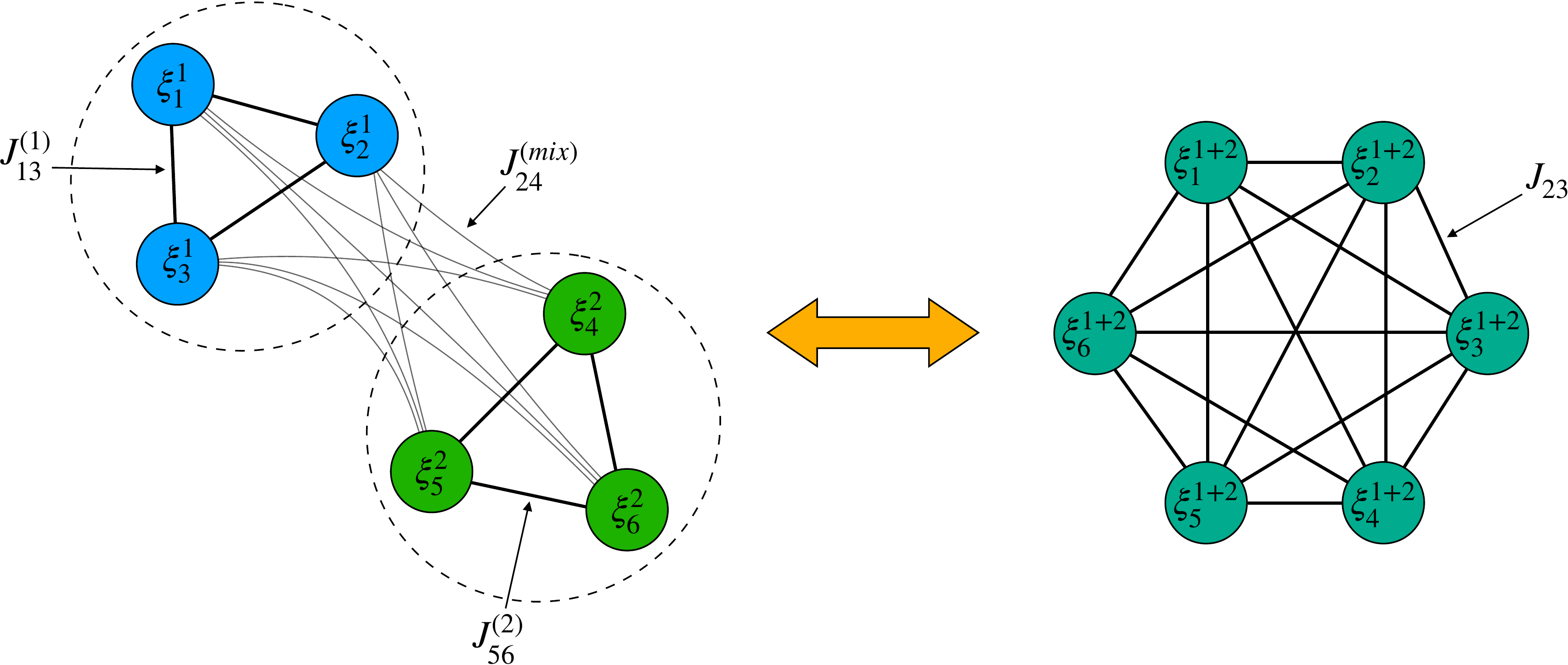}
	\caption{{\bfseries Representation of the classical conditioning within the neural network.} In this picture, on the right the whole system in split into two sub-families such that the neural configuration $\bb \sigma$ is represented as direct sum of two distinct vectors: $\bb \sigma = \bb \xi^{1} + \bb \xi^{2}$ (i.e. half of the neurons account for the pattern $\bb \xi^{1}$ and half of the neurons account for pattern  $\bb \xi^{2}$ the two being uncorrelated, namely $\bb  \xi^{1} \cdot  \bb \xi^{2} = 0$). Accordingly, the synaptic matrix is naturally split into four blocks: $\bb J^{(1)}$ (accounting for the pattern $\bb \xi^{1}$) and $\bb J^{(2)}$ (accounting for the pattern $\bb \xi^{2}$) are the intra-family connections storing information on the single patterns, while the two off-diagonal blocks $\bb J^{(mix)}$ connect neurons of different subfamilies but their synapses grow solely when driven by classical conditioning. See also Figures (\ref{fig1},\ref{Jmix}).}\label{fig:pavlov_bam}. 
\end{figure}

\begin{figure}[tb]
\begin{centering}
\includegraphics[width=0.75\textwidth]{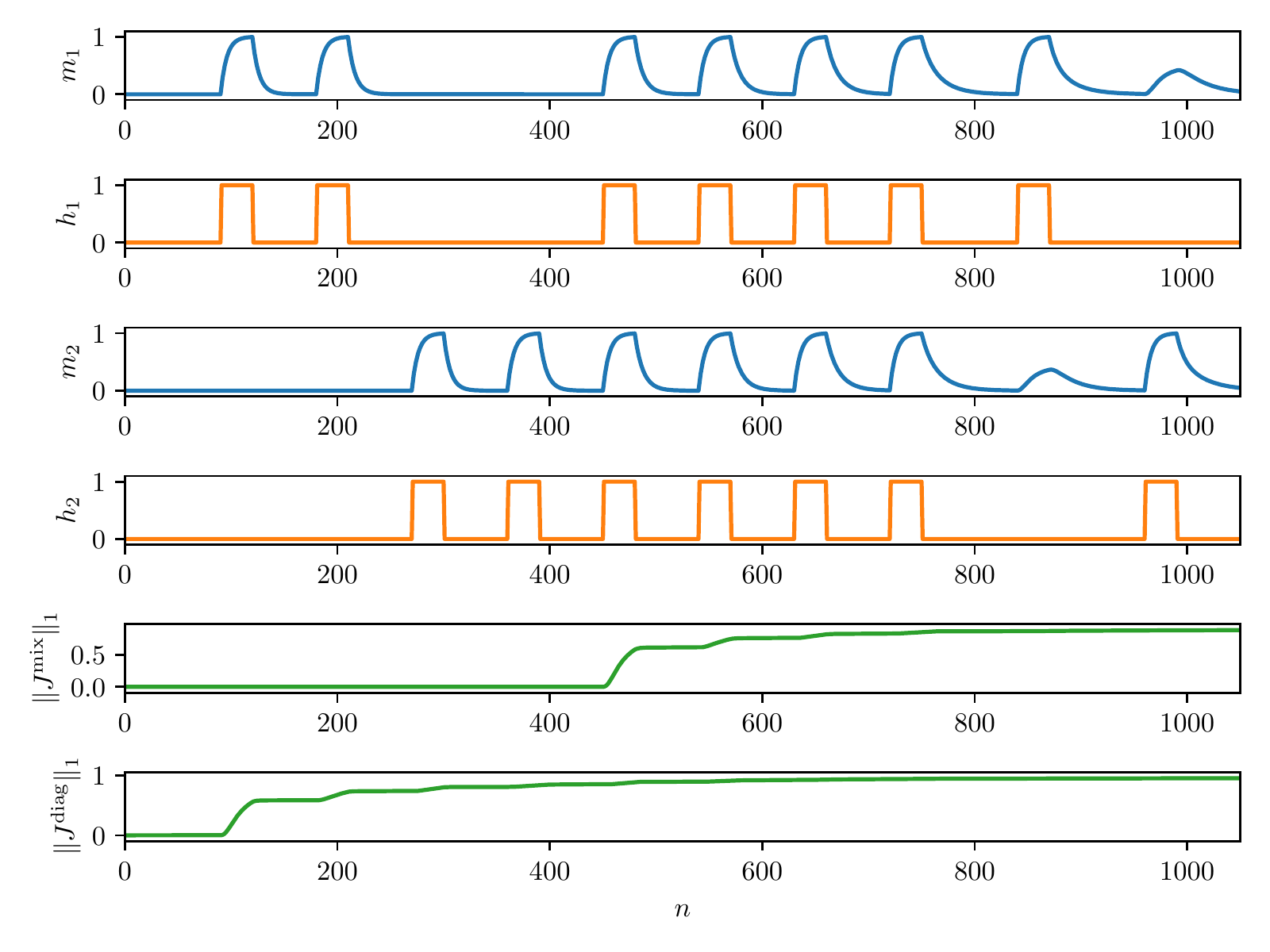}
\par\end{centering}
\caption{\label{fig20}{\bfseries Extended Conditioning via Langevin dynamics on groups of neurons}. The six panels report the evolution of the neural activities through $m_{i}^{(n)}$ (whose definition is a direct generalization of equation \eqref{magsti}),of the stimuli presented to the network through $h_{1,2}$ and of the blocks of the synaptic matrix $\bb J$ through $\left\Vert J^{{\rm mix}}\right\Vert _{1}$ and $\left\Vert J^{{\rm diag}}\right\Vert_{1} $. During the first $250$ iterations the stimuli are presented separately and the network responds properly ($\left\Vert J^{{\rm diag}}\right\Vert_{1} $ raises from zero). In the next $400$ iterations the stimuli are simultaneously presented to the network and $\left\Vert J^{{\rm mix}}\right\Vert _{1}$ raises from zero. As a result when, at $n=850$, we present solely the first stimulus, both the magnetization $m_{1}$ and $m_{2}$ responds. The parameters used for the simulation (to be inserted in equations \eqref{discrete_time_a},\eqref{discrete_time_b}) are $N=20$, $\beta=4$, $u=1$ and $\tau/\tau^{\prime}=0.0004$ and $\delta t=0.16\tau$. }
\end{figure}

\medskip
In order to corroborate numerically this picture we inspect the overlap  between the blocks of the matrix  $\bb J(\bb \xi^{1+2})$ and the time dependent blocks of the  matrix $\bb J^{(n)}$ emerging from our numerical experiment; to this purpose we introduce the (planted) overlaps
\begin{equation}\label{q_planted}
q_{\rm diag}^{(n)}:=\frac{1}{2}\sum_{k=1}^{2}\frac{\sum_{(i,j)\in B_{k}}J_{ij}^{k}J_{ij}^{(n)}}{\sum_{(l,m)\in B_{k}}J_{lm}^{k}J_{lm}^{k}},\quad q_{\rm mix}^{(n)}:=\frac{\sum_{(i,j)\in B_{\rm mix}}J_{ij}^{\,\rm mix}J_{ij}^{(n)}}{\sum_{(l,m)\in B_{\rm mix}}J_{lm}^{\,\rm mix}J_{lm}^{\,\rm mix}}
\end{equation}
Their evolution in time steps is represented Fig.~\ref{fig1}: while the two concepts ($\boldsymbol \xi^1$ and $\boldsymbol \xi^2$) are presented separately (white background), only the blocks related to the single stimuli increases, say blocks $\bb J^{1}$ and $\bb J^{2}$, and accordingly $q_{\rm diag}$ tends to saturate to $1/2$.  On the other hand, while the two concepts are presented simultaneously ($\boldsymbol \xi^1 + \boldsymbol \xi^2$, grey background), the four blocks are all increasing and $q_{\rm diag}$, $q_{\rm mix}$  both saturate to $1$. Otherwise stated, the presentation of $\boldsymbol \xi^{1+2}$ implies conditioning $\bb\xi^{1}$ and $\bb\xi^{2}$, as evidenced by the growing of $q_{\rm mix}$. Eventually the synaptic matrix reaches the form given by \eqref{eq:J_H}.
In Fig.~\ref{Jmix} we show the evolution of the synaptic matrix as the learning dynamics flows: the off-diagonal blocks raise from zero just when both the stimuli are simultaneously presented and eventually all the entries in the matrix take values $\pm 1$. Finally, Fig.~\ref{fig:pavlov_bam} provides a qualitative description of the conditioning phenomenon.\\
The scenario just described is preserved in the case of the patterns involving a different number of neurons, say $p N$ for $\boldsymbol \xi^1$ and $(1-p) N$ for $\boldsymbol \xi^2$, with $p \in (0,1)$, as discussed in Appendix \ref{app:B}.\\
To conclude, we point out that the classical conditions via Langevin dynamics on single couples of neurons can be easily extended to the case of  assemblies of neurons (see Fig.\,\ref{fig20}): the field inserted in eqs.~\eqref{discrete_time_a}-\eqref{discrete_time_a} takes the form given in \eqref{guerrastyle} but now $\bb\xi^{1},\bb\xi^{2}\in\{-1,1\}^{N/2}$.
Again, mimicking Figure \ref{figtest} (obtained with single bits of information) in spirit, in the first part of the experiment the two group of neurons are stimulated separately and each single stimulation session corresponds to an alignment of the related group of neurons and the synapses contained into the off-diagonal blocks remain neutral. In the second part of the experiment the two groups of neurons are stimulated simultaneously, they both react accordingly and the off-diagonal blocks of the synaptic matrix store their correlation. Then, in the third part of the experiment when only one of the two group of neurons is stimulated also the other reacts, confirming that Classical Conditioning took place.
\bigskip

\subsection{Generalized conditioning: integrating multiple signals at once} \label{sec:GC}
In this section we aim to move from the original Pavlovian setting, where only two stimuli are involved, toward more challenging cases where a larger number of stimuli is involved at once (see \emph{e.g.}, \cite{ResWag,Macka}). 
\newline
In particular, we build the following four patterns 
\begin{eqnarray}
\boldsymbol \xi^1&=& (\xi_1^1, \xi_2^1, ..., \xi^1_{\frac{N}{4}},0,0,...,0)\\
\boldsymbol \xi^2&=& (0,0,...,0, \xi^2_{\frac{N}{4}+1},..., \xi^2_{\frac{N}{2}},0,0, ..., 0)\\
\boldsymbol \xi^3&=& (0,0,...,0, \xi^3_{\frac{N}{2}+1}, ,...,\xi^3_{\frac{3N}{4}},0,0, ..., 0)\\
\boldsymbol \xi^4&=& (0,0,...,0, \xi^4_{\frac{3N}{4}+1}, ...,\xi^4_{N}).
\end{eqnarray}
As in the previous section, the patterns have no overlap and each of them stimulates the same amount of neurons, in this case $N/4$. 
The fields inserted in \eqref{discrete_time_a}-\eqref{discrete_time_b} and stemming from these stimuli read as equation \eqref{fieldvar} where, in this experiment, $\boldsymbol \xi^{(n)} =\boldsymbol \xi^{\mu} $ if $n \in \mathcal I_{\mu}$, for $\mu=1,...,4$.

\begin{figure}
\begin{centering}
\includegraphics[width=0.95\columnwidth]{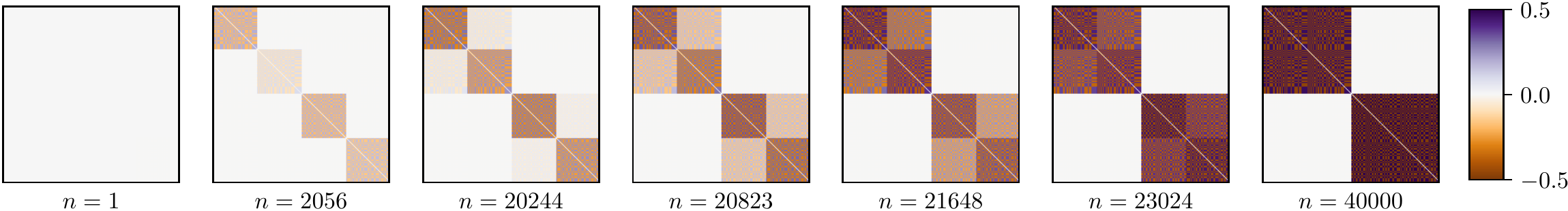}
\par\end{centering}
\caption{\label{1a}\textbf{Evolution of the synaptic matrix when a larger number of stimuli is involved.} Synaptix matrix in several training iterations: for $n\leq 2000$ solely one pattern per time is presented (in particular each pattern is presented consecutively for $n=300$ iterations) and the matrix converges to a block matrix where solely the single patterns $\boldsymbol \xi^{1},\boldsymbol \xi^{2},\boldsymbol \xi^{3},\boldsymbol \xi^{4}$ have been stored). Then, for $n\geq20000$ conditioning take place as the stimuli are now presented simultaneously in couples, in the form $\bb\xi^{1}+\bb\xi^{2}$ or $\bb\xi^{3}+\bb\xi^{4}$: the network correctly stores the correlations among the relative stimuli within each presented couple (hence the relative off-diagonal blocks are increased). The parameters used for the simulation (to be inserted in equations \eqref{discrete_time_a},\eqref{discrete_time_b}) are $N=200$, $\beta=10$, $\tau/\tau^{\prime}=0.005$, $\delta t =0.1\tau$ and the strength of the stimuli is $u=200$.}
\end{figure}

\begin{figure}
\begin{centering}
\includegraphics[width=0.95\columnwidth]{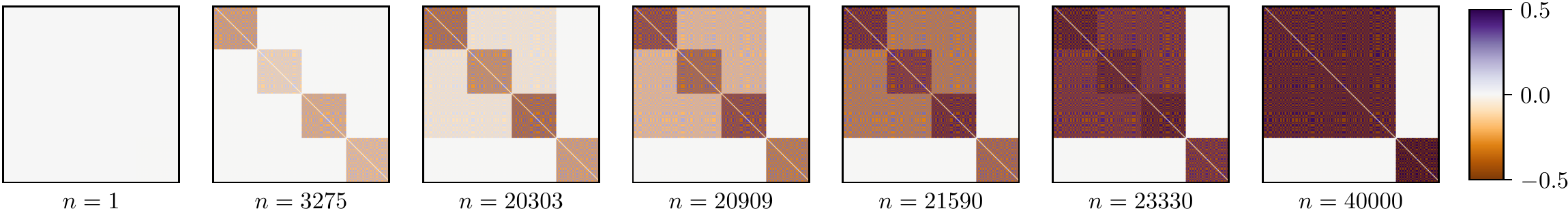}
\par\end{centering}
\caption{\label{2a}\textbf{Evolution of the synaptic matrix.} Synaptix matrix in the various training iterations $n$ of the second experiment: for $n\leq20000$ solely one pattern per time is presented and the matrix converges to a block matrix where only the single patterns $\bb\xi^{1},\bb\xi^{2},\bb\xi^{3},\bb\xi^{4}$ have been stored). Then, for $n\geq 20000$  conditioning take place as the stimuli are now presented simultaneously in couples, in the form $\bb\xi^{1}+\bb\xi^{2}+\bb\xi^{3}$ and, separately, $\bb\xi^{4}$: the network correctly stores the correlations among the relative three stimuli that are jointly presented  (hence the relative off-diagonal blocks are increased). The parameters of the simulation are the same as Fig.\,\ref{1a}. }
\end{figure}

The training session is scheduled analogous to the one outlined in the previous section for Classical Conditioning, as we briefly recall. First, the patterns are presented and learnt separately and $\bb J^{(n)}$ exhibits a block-shape where only diagonal blocks are non-null; next, we present two or more of them simultaneously -- note that in this case we have the freedom to group patterns to be presented simultaneously and here we consider the case where they are presented in couples like $\boldsymbol \xi^{1+2} = \boldsymbol \xi^1 + \boldsymbol \xi^2, ~ \boldsymbol \xi^{3+4} = \boldsymbol \xi^3 + \boldsymbol \xi^4$ and the case we they are presented in two inhomogeneous groups like $\boldsymbol \xi^{1+2+3} = \boldsymbol \xi^{1} + \boldsymbol \xi^{2} + \bb \xi^3, ~ \bb \xi^4$ -- and we can check that here the non-null diagonal blocks of $\bb J^{(n)}$ mirrors the combination of patterns presented; finally, in the third tranche of the training session, we present all the four patterns simultaneously to inspect if the missing correlations (among patterns beloning to different groups) now appear in the synaptic matrix and indeed this is the case, see Figs.~\ref{1a}, \ref{2a}.
\newline


\subsection{Convergence of two-scale Langevin dynamics to the Hebbian synaptic matrix} \label{sec:Hebb}

\begin{figure}
	\centering
	\includegraphics[width=\textwidth]{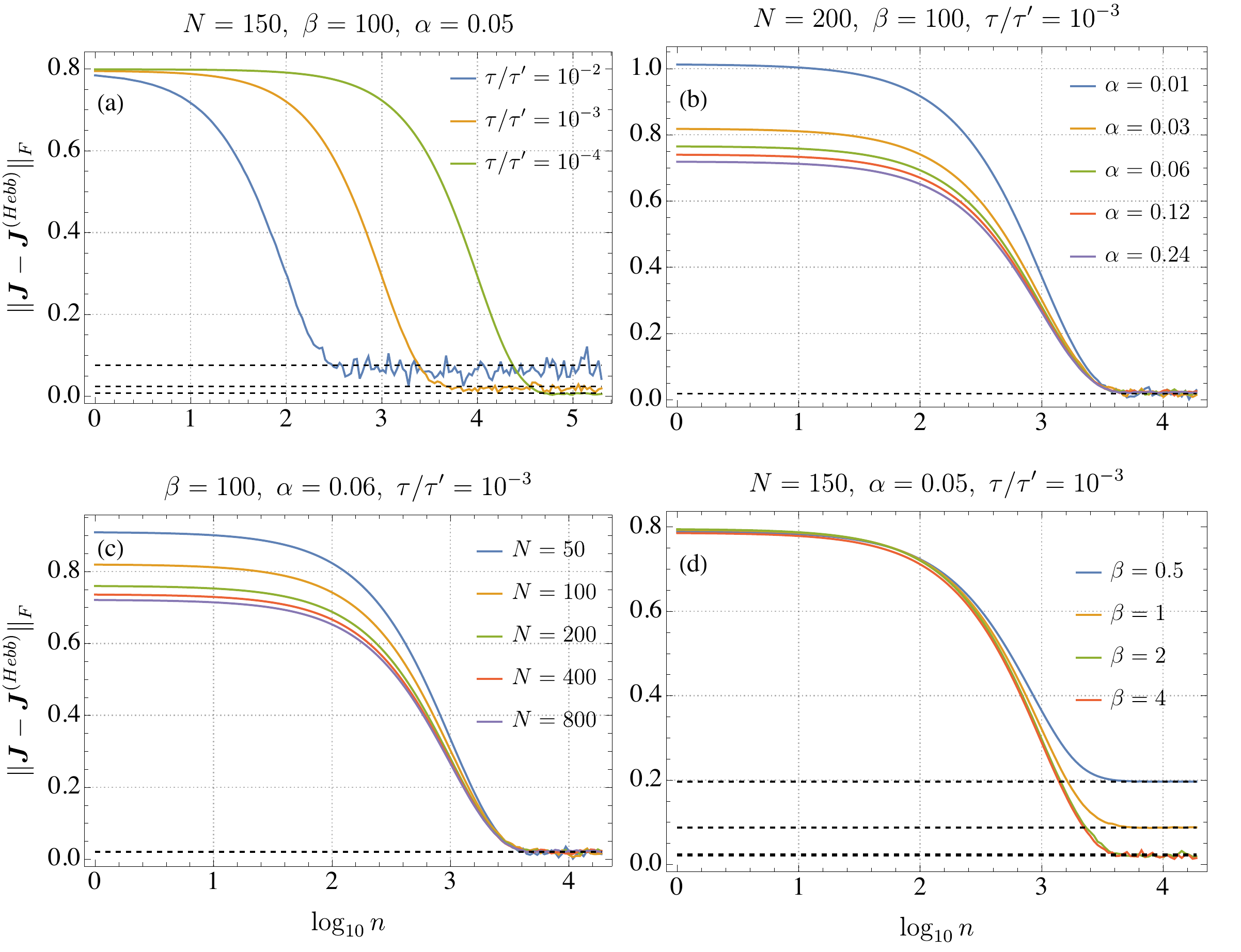}
\caption{{\bfseries Results for the Langevin dynamics with multiple external stimuli: convergence to the Hebbian kernel.} The figure shows the results of the Langevin dynamics for various setting of the external parameters. The curves refer to the Frobenius norm of the difference between the synaptic matrix and the corresponding Hebbian kernel as a function of the evolution time. In panel a, we fix  the network size $N=150$, the storage capacity $\alpha=0.05$ and the thermal noise level $\beta =100$ and vary the ratio $\tau/\tau' =10^{-2}$, $10 ^{-3}$ and $10^{-4}$. In panel b, we fix the network size $N=200$ and the ratio $\tau/\tau'=10^{-3}$ and vary the storage capacity $\alpha=0.01, \ 0.03, \ 0.06, \ 0.12, \ 0.24$. In panel c, we fix the storage capacity $\alpha =0.06$ and the ratio $\tau/\tau'=10^{-3}$ and vary $N=50,\ 100, \ 200, \ 400, \ 800$. In panel d, we fix the network size $N=150$, the storage capacity $\alpha=0.05$ and the ratio $\tau/\tau'=10^{-3}$ and vary the thermal noise $\beta = 0.5, \ 1, \ 2, \ 4$ (the fluctuations are now computed according to Eq. \eqref{eq:beta_fluct}). Each pattern is presented to the network as external field, the duration of the stimulus being restricted to a single Langevin step. The pattern is selected with uniform probability in the index set $\{1,\dots,K\}$ and the procedure is repeated for $2\cdot 10^4$ evolution steps. Moreover in all the simulations the strength of the stimuli is equal to $u=200$.}\label{converg}
\end{figure}

\begin{figure}
	\begin{centering}
		\includegraphics[width=0.75\textwidth]{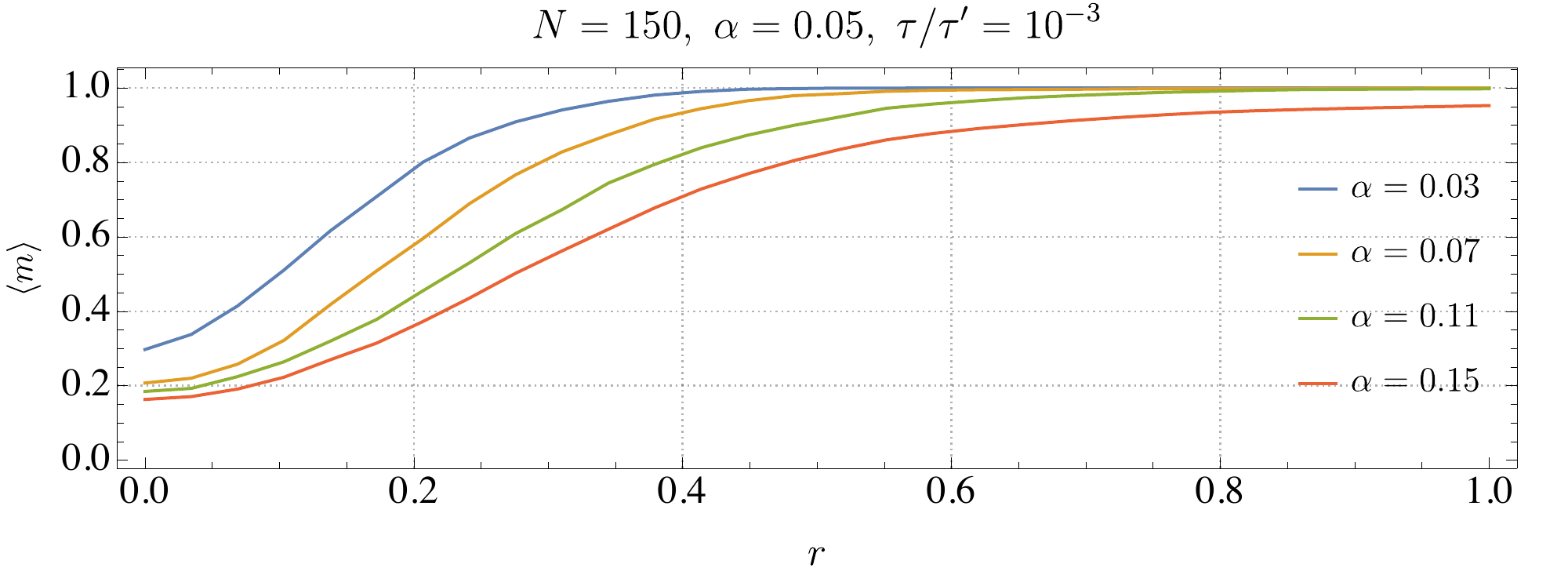}
		\par\end{centering}
	\caption{\label{retrieval} {\bfseries Hebbian-like retrieval capabilities of the network after training.} For each of the $K$ patterns stored in the network, we generate $M=100$ noisy versions of them by flipping randomly a percentage of bits equals to $\frac{1-r}{2}\times100$ with $r\in(0,1]$. These examples are used as initial condition for the dynamics, then we let the system evolves for $5$ iterations with the dynamics prescribed by equation \eqref{eq:sigma_rad}, at fixed synaptic matrix $\bb J$ and in absence of stimuli $h_{i}=0$. We evaluate the mean Mattis magnetization: $m_{\mu}=\frac{1}{M}\sum_{a=1}^{M}\bb \sigma^{\rm final}_{a}\cdot\bb \xi^{\mu}/N$ between the $M$ final neural configurations $\bb\sigma^{\rm final}_{a}$ and the pattern $\mu$. In particular, $\bb J$ is the matrix obtained in the end of the training session described in panel a, orange curve. We repeat this algorithm for all the patterns
	and we average over the values of $m_{\mu}$ obtained: $\langle m\rangle=\frac{1}{K}\sum_{\mu=1}^{K}m_{\mu}$. The parameters of the simulations are $N=150$ and $\beta =100$. Note that for $\alpha>\alpha_c \sim 0.14$ the magnetization does not saturate any longer, in agreement with AGS theory.}
\end{figure}

Up to now we dealt with a relatively small number of patterns, each involving a subset of the neurons making up the system, and we showed that -beyond recovering Classical Conditioning- when they are combined the synaptic matrix tends to reach a Hebbian structure. In this section we directly consider patterns that involve the whole set of neurons, namely they occur as $\boldsymbol \xi^{\mu} = (\xi_1^{\mu}, ..., \xi_N^{\mu})$, with $\mu=1,...,K$, where $K$ can be very large, possibly scaling with the volume $N$.  
As we are going to prove, as long as pattern entries are Rademacher, that is
$$
P(\xi^\mu_i = \pm1)=\frac12 ,\qquad \forall i=1,\dots,N \text{ and } \mu=1,\dots,K,
$$
we randomly choose one of the $K$ patterns as external stimulus of the network at each iteration, the temperature is sufficiently high $\beta \gg 1$ and the ratio between the time scales is low ($\tau/\tau^{\prime}\ll 1$),
the long-term limit of the evolution provided by \eqref{discrete_time_a}-\eqref{discrete_time_b} generates the Hebbian matrix
\begin{equation} \label{eq:J_hebb}
J_{ij}^{(Hebb)}=\frac1K \sum_{\mu=1}^K \xi^\mu_i \xi^\mu_j,
\end{equation}
thus the AGS statistical mechanical theory is asymptotically recovered as it should. To be more precise, in Appendix \ref{sec:fluct} we dimostrate that, in the end of the training, the synaptic matrix converges to the temporal mean of the stored patterns plus a fluctuation (vanishing in the limit $\beta\to\infty$ and $\tau/\tau^{\prime}\to0$) and, since in these simulations at each time step $n$ the stimulus can be $\bb \xi^{\mu}$ with probability $1/K$, the temporal mean of the stored patterns coincides with the Hebbian prescription for the synaptic matrix as given in \eqref{eq:J_hebb}.\\
For the sake of simplicity, we set $\delta t=\tau$ in equations \eqref{discrete_time_a},\eqref{discrete_time_b}: this means that the neurons align themselves with the stimulus instantly thus we don't need to present the same stimulus for several iterations, just one iteration is sufficient to produce the alignment of the neurons\footnote{In Sec.\,\ref{sec:IV} we set $\delta t =c\tau$ with $c\in(0,1)$ because we wanted to observe the neural reaction to an external stimulus and appreciate the transient effects of the presence of an appropriate time scale for neurons (see for example Fig.\ref{figtest}). Now we want to focus only on the dynamics of the synaptic matrix, hence the decision to set $c=1$.}.
In a nutshell, the learning procedure consists in presenting these $K$ patterns as external stimulus to the network, so that the relaxation ends up in a configuration in which the coupling matrix retains the information.  To do this, we randomly choose one of the patterns, say $\bb \xi^{\bar \mu}$, and set the external field to $\bb h = \bb \xi^{\bar \mu}$ for a single evolution step; this procedure is repeated as the time $t$ goes by. 
In Fig. \ref{converg}, we reported the distance, computed according to the Frobenius norm
$$
\lVert A \rVert _F := \sqrt{\frac1{N^2}\sum_{ij=1}^N A_{ij}^2},
$$
of the resulting coupling matrix $\bb J^{(n)}$ at each time step w.r.t. the Hebbian kernel eq.~\eqref{eq:J_hebb} for different values of the ratio $\tau/\tau'$ (panel a) and parameters $\alpha=K/N$ (panel b), $N$ (panel c) and $\beta$ (panel d). First, we notice that if the noise is low enough (\emph{i.e.} $\beta\gg 1$) and if the ratio $\tau/\tau^{\prime}$ is low enough, the dynamics of the coupling matrix converges to the Hebbian kernel. As a matter of fact, there is always a reconstruction error of the Hebbian kernel, in fact, the Frobenius norm of the difference between the synaptic matrix and the Hebbian kernel has an asymptotic value which, in general, is different from zero and depends on $\tau/\tau^{\prime}$ and $\beta$ (see panel a,d). In Appendix \ref{sec:fluct} we study the fluctuations of the coupling matrix $\bb J$, resulting from the Langevin dynamics around the Hebbian fixed point given in \eqref{eq:J_hebb}. We find that in the case $\beta\gg1$ the fluctuation is vanishing like ${(\tau/\tau')}^{1/2}$ (see equation \eqref{eq:fluct}) and perfectly reproduces the asymptotic behavior of the Frobenius norm of the difference between the synaptic matrix and the corresponding Hebbian kernel (see panel a, dashed line). Whereas, if $\beta\sim 1$ we have to take into account also the contribution of the fast noise to the fluctuation around the Hebbian kernel and this is given by a more general formula (see equation \eqref{eq:beta_fluct}), also in this case we have a perfect agreement between the theoretical estimations and the numerical simulations (panel d, dashed line). Finally, in the panels b and c, we observe that the fluctuation around the Hebbian kernel does depend nor on the size of the system $N$ and neither on the parameter $\alpha=K/N$ as expected from the theoretical estimation; this implies that the algorithm is robust in performing the storage of the patterns in the synaptic matrix. Moreover, for values of the storage capacity larger than $\alpha_c \sim 0.14$  the magnetization does not saturate to one any longer in full agreement with the statistical mechanical limiting description provided by AGS theory \cite{amit1985}, see Fig.~\ref{retrieval}.

\subsection{Persistent retrieval, obsessions and unlearning} \label{sec:obs}

\begin{figure}[h!]
	\centering
	\includegraphics[width=0.65\textwidth]{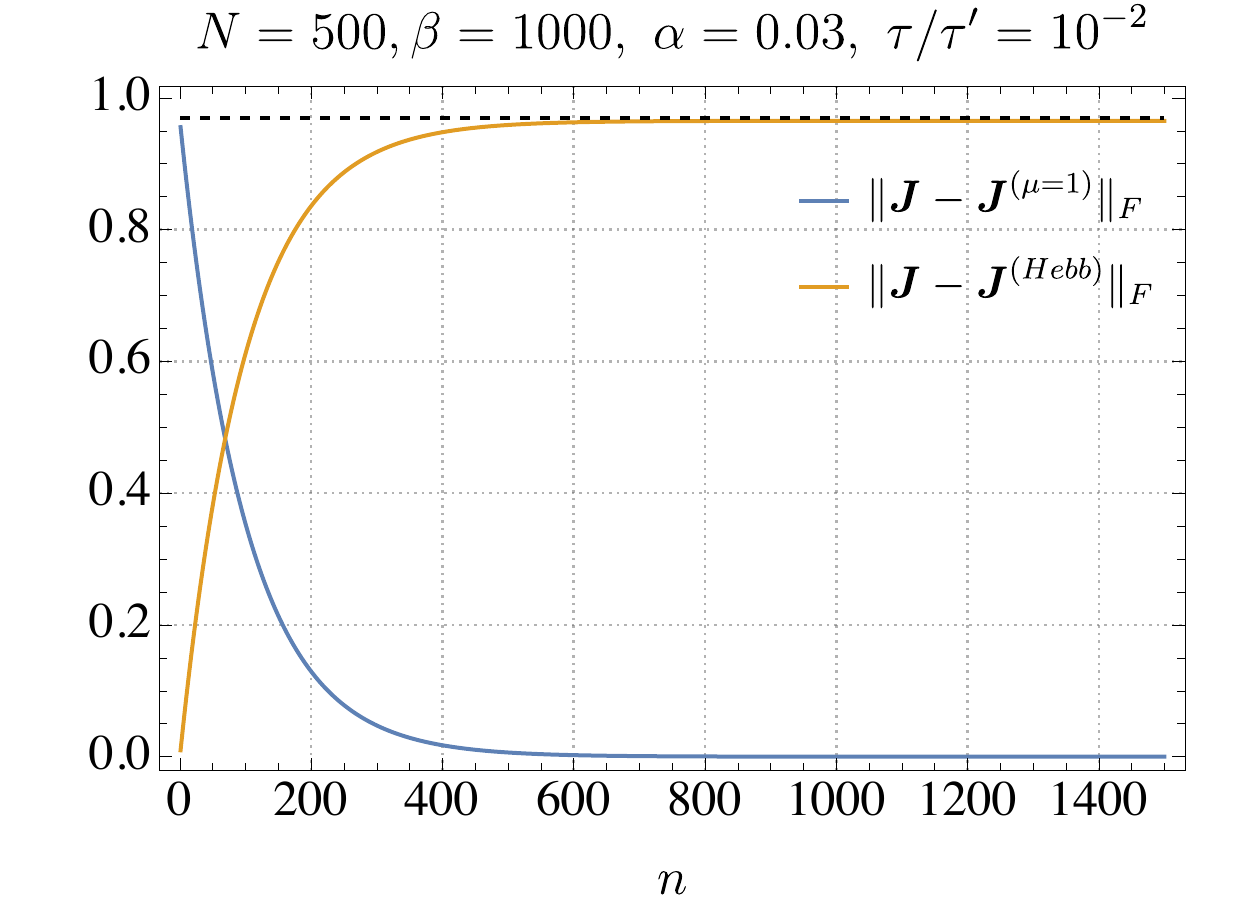}
	\caption{{\bfseries Persistent retrieval of the same pattern as natural memory erasure.}  The evolution of the Frobenius norms $||J-J^{Hebb}||_{F}$ and $|J-J^{\mu=1}||_{F}$  when the first pattern is persistently retrieved proves that -after enough time- the whole memory reduces to the first pattern, all the other being erased by this peristency, making the network hooked. The simulation reported in the plot has been carried by keeping $\tau/\tau' = 10^{-2},\ \alpha=0.03,\ \beta=10^3,\ N=500$; the black dashed line is the constant function $\sqrt{1-\frac{1}{K}}\approx0.96$.}\label{fig:stupid}
\end{figure}

Beyond the outlined structural similarities between Pavlov's and Hebb's representations of learning, the intrinsically dynamical process underlying Pavlovian association mechanism highglights a phenomenon that can not be captured by the statistical mechanical approach underlying Hebbian storage \cite{amit1989,CKS}:  if the network gets stuck in retrieving persistently the same pattern (\emph{i.e.}, an {\em obsession}), each retrieval reinforces always the same minimum in the memory landscape up to the point that the latter destroyes all the other memories and solely the obsessive pattern remains stored. In Fig.~\ref{fig:stupid} we report results of the following numerical experiment that confirms such a scenario: starting from a perfect Hebbian kernel $J^{(Hebb)}$,  we let the nework retrieve always the first pattern (\emph{i.e.}, the field is persistently $\bb h=\bb \xi^1$) and we evaluate, iteratively in $n$, the evolution both of  $||J^{(n)}-J^{(Hebb)}||_{F}$ and  $||J^{(n)}-J^{(\mu=1)}||_{F}$, where $J^{(\mu=1)}=\xi_i^1\xi_j^1$. Remarkably, in the long time limit, the synaptic matrix collapses onto $\xi^1$ (i.e. $\lim_{n \to \infty} J^{(n)} = J^{(\mu=1)}$), while $||J^{(n)}-J^{(Hebb)}||_F \to  \sqrt{1-\frac{1}{K}}$ as expected.
\newline
It is important to put emphasis on this phenomenon because  in classical learning theories  pattern recognition can not alter the memory landscape, but this is not the case here (further this can also be seen as a technique for forgetting alternative to pruning \cite{Pruning} or dreaming \cite{FAB-NN2019} mechanisms).

\section{Conclusions}\label{sec:V}

A number of observations can be drawn from this research.
\newline
The first is that proving that a two-scale Langevin dynamics gives rise spontaneously to the Palvov mechanism and that the latter, in the long term limit, converges to the Hebbian synaptic matrix was a (somehow expected but missing) bridge between multi-scale stochastic processes and statistical mechanics of neural networks. Even by this new perspective, in particular by its generality, we are prone to think that these kinds of information processing mechanisms are quite spread in Nature and actually not confined at all within the neural world. Further, by the same perspective outlined in this paper, there is no excitatory or inhibitory learning, rather there is just statistical learning (then patterns with  positive correlations would be reciprocally excitatory viceversa for anti-correlated patterns, but the whole process of learning is entirely statistically driven), likewise we do not distinguish with this level of modeling between primary and secondary stimuli (each pattern plays as a conditioner for each other as long as we work in a random setting).
\newline
A remarkable accordance between broad heuristic evidence on Classical Conditioning and our formulation of this phenomenon is that, driven by the Hebbian paradigm (\emph{i.e.}, by looking at a neuron as a temporal integrator circuit), the only way Classical Conditioning can take place is by presenting two stimuli acting on different areas (\emph{i.e.} different neurons) and then correlations among the stimuli will result in growth of synapses connecting these (previously uncorrelated) neurons or cliques of neurons\footnote{This also partially explain why previous approach of this kind were absent in the Literature, as the reference Hebbian network to deal with can not be those of AGS but must be the multitasking ones, whose origin is much more recent \cite{Multitasking,Hierarchical,ScaleFree}.}.
\newline
Further, a remark on Generalized Conditioning, closer to Rescorla-Wagner \cite{ResWag} and Mackintosh \cite{Macka} researches in spirit: by connecting Pavlov mechanisms to Hebbian learning, we also tacitely proved the existence of phase transitions as the memory  storage is made to vary. Some intrinsic discrepancies among the expected effectiveness of a conditioning (or  its expected lacking for some particular conditioning mechanisms) and the reality of the experiments  (that are still today a puzzle in the field \cite{ReviewP}) could be explained by the fact that while making the experiments eventually the critical capacity has been reach, hence saturation mechanisms, retrieval, etc. all these responses sensibly diminish (as we have shown in Figure \ref{retrieval}) 
\newline
Finally, as Pavlov's learning rule is intrinsically dynamics, there are phenomena that it describes but that can not be captured by a statistical mechanical picture: by the latter patterns (e.g. in the Hopfield model) have their stable basins of attraction and these are static  if no further storing takes place (namely the network can retrieve persistently the same pattern and this operation does not alter the landscape where stored memories lie): solely learning new patterns alters the memory landscape.  From the Pavlov perspective, instead,  as shown in Figure \ref{fig:stupid}, retrieval affects the amplitude and stability of minima and retrieving persistently the same pattern makes its  basin of attraction more and more pronounced w.r.t. the basins of the other patterns (up to the point that, by constantly retrieving the same pattern -- modeling obsessions -- the whole memory collapses to solely that information), hence, {\em forgetting} within the Pavlov scheme can be achieved without network's pruning \cite{Pruning,pruning2} or removal via sleeping induced mechanisms \cite{FAB-NN2019,last-dream}.

\appendix
\section{Langevin dynamics and Hebb relaxation from Gaussian prior for the synapses} \label{ssec:gauss}
In this case, the space of interaction matrices is endowed with the probability measure
$$
dP(\bb J)= \prod_{i<j} \frac{dJ_{ij}}{\sqrt{2\pi}}\exp\big(- J_{ij}^2/2\big).
$$
Again, we recast the partition function \eqref{eq:Z_0} with the addition of source terms $\bb t$ and $\bb T$ for, respectively, the neurons and the weights:
\begin{equation}
{Z}_{N}(\beta, \bb t, \bb T)=
\sum_{\bb\sigma\in \Sigma_N}\int\big(\prod_{i<j}\frac{dJ_{ij}}{\sqrt{2\pi}}\big)\exp\Big(-\frac{1}{2}\sum_{i<j}J_{ij}^{2}+\beta\sum_{i<j}J_{ij}\sigma_{i}\sigma_{j}+\beta u\sum_{i}h_{i}\sigma_{i}+\sum_{i}t_{i}\sigma_{i}+\sum_{i<j}T_{ij}J_{ij}\Big).\label{p(s)-1}
\end{equation}
%
In fact, performing the integration w.r.t the prior distribution $P(\bb J)$, we have
\begin{equation}
\begin{split} &{Z}_{N}(\beta, \bb t, \bb T)=\sum_{\bb\sigma\in\Sigma_N} \exp\Big(\beta u\sum_{i}h_{i}\sigma_{i}+\sum_{i}t_{i}\sigma_{i}+\frac{1}{2}\sum_{i<j}\left(\beta\sigma_{i}\sigma_{j}+T_{ij}\right)^{2}\Big),
\end{split}
\end{equation}
and the expectation value of the synapses is evaluated in terms of the derivative w.r.t. the source, {\it i.e.}
\begin{equation}
\langle J_{ij}\rangle=\frac{\partial\log {{Z}_{N}(\beta, \bb t, \bb T)}}{\partial T_{ij}}\Big\vert_{\bb T=0, \bb t =0}=\beta\langle\sigma_{i}\sigma_{j}\rangle\underset{m.f.}{\approx}\beta\langle\sigma_{i}\rangle\langle\sigma_{j}\rangle\label{one}.
\end{equation}
For the neural correlation functions, we proceed in analogous manner and compute
\begin{equation}
\nonumber
\begin{split} &
 \sum_{\sigma_{i}=\pm1}P(\bb \sigma,\bb J)=
 \frac{1}{{Z}_{N}(\beta, \bb t, \bb T)}\ 2\cosh\big(\beta\sum_{j\neq i}J_{i j}\sigma_{j}+u\beta h_{i}+t_{i}\big)\exp\Big(\beta
 \sum_{\underset{k<j}{k,j\neq i}}
 J_{kj}\sigma_{k}\sigma_{j}+\sum_{k\neq i}\left(t_{k}+u \beta h_{k}\right)\sigma_{k}\Big).
\end{split}
\end{equation}
The expectation value of the neural activity $\sigma_i$ is therefore obtained as
\begin{equation} \nonumber
\langle\sigma_{i}\rangle=\frac{\partial\log {{Z}_{N}(\beta, \bb t, \bb T)}}{\partial t_i}\Big\vert_{\bb T=0, \bb t =0}=\langle\tanh\big(\beta\sum_{j\neq1}J_{i j}\sigma_{j}+u\beta h_{i}\big)\rangle\underset{m.f.}{\approx}\tanh\big(\beta\sum_{j\neq i}\langle J_{ij}\rangle\langle\sigma_{j}\rangle+u\beta h_{i}\big),
\end{equation}
where we used again the mean-field assumption. Since all neural indexing is equivalent, the same equality holds \emph{mutatis mutandis} for each $i=1,\dots,N$. 
\newline
With these results, we can set up a Langevin-like dynamics governed by the following evolutive equations:
\begin{eqnarray}
\label{eq:ps1}
	\dot{\langle{\sigma_{i}}\rangle}&=&-\frac1\tau\langle\sigma_{i}\rangle+\frac1\tau\tanh\big(\beta\sum_{j\neq i}^{N}\langle J_{ij}\rangle\langle\sigma_{j}\rangle+u\beta h_{i}\big),\\
	\label{eq:ps2}
	\dot{\langle{J_{ij}}\rangle}&=&-\frac1{\tau'}\langle J_{ij}\rangle+\frac\beta{\tau'}\langle\sigma_{i}\rangle\langle\sigma_{j}\rangle,
\end{eqnarray}
where $\tau$ and $\tau'$ are, respectively, the neural and synaptic time-scales. 
\newline
It is worth pointing out that the long-term relaxation of eq. (\ref{eq:ps2}) -i.e.  achieved by imposing  $\dot{\langle{J_{ij}}\rangle}=0$ and addressed by the relative statistical mechanical picture of the network-  prescribes $\langle J_{ij} \rangle = \beta\langle \sigma_i \rangle \langle \sigma_j \rangle$ such that, by applying two single-bit stimuli $\xi_i$ and $\xi_j$ on the two neurons $\sigma_i$ and $\sigma_j$ (such that  $\sigma_i = \xi_i$ and $\sigma_j = \xi_j$) we recover the Hebbian structure for the synaptic matrix, namely
$$
J_{ij} = \beta \xi_i \xi_j,
$$
where the noise term $\beta$ simply shifts the critical intensity of the stimuli for learning to take place. The difference between this scenario and the one obtained for Rademacher priors, see \eqref{eq:JJ}, is the role of the thermal noise in the evolution of couplings. Indeed, in the Rademacher case, the level of thermal noise enters as $\tanh \beta$, which is limited as it should.

\section{Classical conditioning for patterns of different length} \label{app:B}
\begin{figure}
\begin{centering}
\includegraphics[width=0.9\columnwidth]{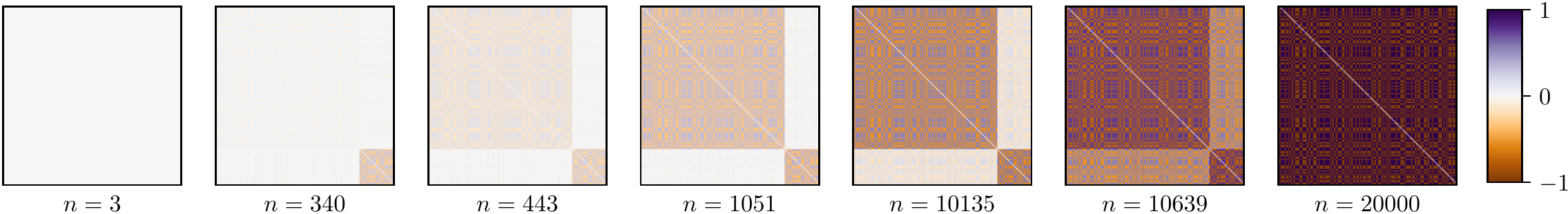}
\par\end{centering}
\caption{\label{abbastardo2}Evolution of the synaptic matrix $J$  while learning single patterns for $n\leq10000$ and when relative conditioning takes place for $n> 10000$. We can appreciate how, while solely learning of independent pattern is present, off-diagonal blocks stay null, viceversa, they start to raise as the concepts are simultaneously presented and conditioning starts. In the simulation $p=0.8$.}
\end{figure}

Now  we still keep the assumption that we deal solely with two signals (the latter will be removed later on in this subsection), but we relax the constraint that the lenght (in bits) of the patterns has to be the same for $\xi^{1}$ and $\xi^{2}$: in general,  patterns can be arranged such that
\begin{eqnarray}
\boldsymbol \xi^1&=& (\xi_1^1, \xi_2^1, ..., \xi^1_{Np},0,0,...,0)\\
\boldsymbol \xi^2&=& (0,0,...,0, \xi^2_{Np+1}, \xi^2_{Np+2}, ..., \xi^2_N).
\end{eqnarray}
where $p\in(0,1)$, and thus we question if, in this generalized setting, Classical Conditioning is preserved. 
\newline
Once having presented them separately to the network and after enough persistency (for long enough learning dynamics such that they can be stored), the resulting synaptic matrix reads as 

\begin{equation}
J_{ij}(\{\boldsymbol \xi^1, \boldsymbol \xi^2 \})=\frac{1}{2}\times\begin{cases}
\xi^1_{i}\xi^1_{j} & (i,j)\in B_{1}\\
\xi^2_{i} \xi^2_{j} & (i,j)\in B_{2}\\
~0 & (i,j)\vee (j,i)\in B_{mix}\\
\end{cases}.
\end{equation}
where
$
B_{1}=\left[1,pN\right]\times\left[1,pN\right],\, B_{2}=\left[pN+1,N\right]\times\left[pN+1,N\right],\, B_{mix}=\left[1,pN\right]\times\left[pN+1,N\right]
$
thus, as expected, $\bb J(\{\boldsymbol \xi^1, \boldsymbol \xi^2 \})$ is still a block-matrix but it has obviously a rectangular shape now.
\begin{equation}\label{scorrelata}
\bb J(\{\boldsymbol \xi^1, \boldsymbol \xi^2 \})=\frac{1}{2}\times\left(\begin{array}{cc}
\boldsymbol{J}^{1} & \boldsymbol{0}\\
\boldsymbol{0} & \boldsymbol{J}^{2}
\end{array}\right),
\end{equation}
where  $\boldsymbol{J}^{1}\in\mathbb{R}^{Np\times Np}$, $\boldsymbol{J}^{2}\in\mathbb{R}^{N(1-p)\times N(1-p)}$,$\boldsymbol{0}\in\mathbb{R}^{N(1-p)\times Np}$.
As it can be appreciated by a glance at Figure \ref{abbastardo2}, Classical Conditioning is robust against this kind of perturbation: this is remarkable because, roughly speaking, it states that -no matter the information content of a pattern, it can still play as a conditioner. Moreover, the behavior of the overlaps $q_{\rm diag},q_{\rm mix}$ versus the time step turns out to be exactly the same as that of Fig. \ref{fig1}.

\section{\label{dimostrazione}Analytical estimate of fluctuations around the Hebbian fixed point} \label{sec:fluct}
In this appendix we quantify the fluctuations of the coupling matrix $\bb J$, resulting from the Langevin dynamics, w.r.t. the Hebbian kernel. To this aim, we go back to the continuum version of the dynamical equations and note that if the external stimulus is strong enough ($u\gg1$) and  the temperature is sufficiently low ($\beta\gg1$) equation \eqref{eq:sigma_rad} can be approximated as follows:
\begin{equation}\label{sigris}
\dot{\langle{\sigma_{i}}\rangle}=-\langle\sigma_{i}\rangle+\tanh\big(\beta\sum_{j\neq i}^{N}\langle J_{ij}\rangle\langle\sigma_{j}\rangle+u\beta \xi_{i}(t)\big)\approx-\langle\sigma_{j}\rangle+\frac{1}{\tau}\xi_{i}(t)
\end{equation}
where we have used $h_{i}=\xi_{i}$.
To lighten the notation let us pose $\langle J_{ij}\rangle=J_{ij}$ and $\langle \sigma_{i}\rangle=\sigma_{i}$.
Equation \eqref{sigris} can be easily solved for $\sigma_{i}$:
\begin{equation}
\sigma_{i}(t)=\frac{1}{\tau}\int_{0}^{t}\xi_{i}(t^{\prime})e^{-\frac{t-t^{\prime}}{\tau}}dt^{\prime}.
\end{equation}
The last equation can be rewritten as 
\begin{equation}\label{quasi}
\sigma_{i}(t)=\frac{\int_{0}^{t}\xi_{i}(t^{\prime})e^{-\frac{t-t^{\prime}}{\tau}}dt^{\prime}}{\int_{0}^{t}e^{-\frac{t-t^{\prime}}{\tau}}dt^{\prime}}(1-e^{-\frac{t}{\tau}}).
\end{equation}
and, at this point, it is easy to see that if the same stimulus is presented for $t\gg\tau $ then equation \eqref{quasi} can be approximated as follows:
\begin{equation}\label{apps}
	\sigma_{i}(t)\approx\xi_{i}(t).
\end{equation}
By inserting \eqref{apps} into the dynamical equation for $J_{ij}$, i.e. $\eqref{eq:sigma_rad}$, we get
\begin{equation}
\dot{J}_{ij}=-\frac{1}{\tau^{\prime}}J_{ij}(t)+\frac{1}{\tau^{\prime}}\xi_{i}(t)\xi_{j}(t).
\end{equation}
Finally, by mapping $t\to \tau t$, the dynamical equations of the synaptic matrix becomes
\begin{equation}\label{Jagg_t}
\dot{J}_{ij}(t)=-\frac{\tau}{\tau^{\prime}}J_{ij}(t)+\frac{\tau}{\tau^{\prime}}\xi_{i}(t)\xi_{j}(t).
\end{equation}
Let us making the following ansatz for $J_{ij}$:
\begin{equation}\label{hebb_dim}
J_{ij}(t)=\frac{1}{t}\int_{0}^{t}\xi_{i}(t^{\prime})\xi_{j}(t^{\prime})dt^{\prime}+\delta_{ij}(t),
\end{equation}
meaning that we separate two components in the coupling matrix, {\it i.e.} a linear integration of the external stimuli and a fluctuation contribution (the $\bb\delta$ matrix). With this assumption, Eq. \eqref{Jagg_t} becomes an evolutive equation for the fluctuation term:
\begin{equation}
\dot{\delta}_{ij}(t)+\frac{1}{t}\xi_{i}(t)\xi_{j}(t)-\frac{1}{t^{2}}\int_{0}^{t}\xi_{i}(t^{\prime})\xi_{j}(t^{\prime})dt^{\prime}=-\frac{\tau}{\tau'}\frac{1}{t}\int_{0}^{t}\xi_{i}(t^{\prime})\xi_{j}(t^{\prime})dt^{\prime}-\frac{\tau}{\tau'}\delta_{ij}(t)+\xi_{i}(t)\xi_{j}(t).
\end{equation}
In the large $t$ limit, the first two terms on the l.h.s. vanish, while the integration parts reduce to the usual Hebbian kernel (since, in this case, the temporal average is equivalent to the expectation value in the pattern space because of the ergodic hypothesis). This means that, in this limit, we can rewrite the entire equation in a simpler form
\begin{equation}
\dot{\delta}_{ij}+\frac{\tau}{\tau'}\delta_{ij}(t)=-\frac{\tau}{\tau^{\prime}}J_{ij}^{(Hebb)}+\frac{\tau}{\tau^{\prime}}\xi_{i}(t)\xi_{j}(t).
\end{equation}
By squaring and integrating in $t$ the last equation we get
\begin{equation}
\int_{0}^{\tau^{\prime}}\left[\dot{\delta}_{ij}^{2}+2\dot{\delta}_{ij}\frac{\delta_{ij}(t)\tau}{\tau^{\prime}}+\left(\frac{\delta_{ij}(t)\tau}{\tau^{\prime}}\right)^{2}\right]dt=\int_{0}^{\tau^{\prime}}\left[-\frac{\tau}{\tau^{\prime}}J_{ij}^{(Hebb)}+\frac{\tau}{\tau^{\prime}}\xi_{i}(t)\xi_{j}(t)\right]^{2}dt.\label{integ}
\end{equation}
Since $\bb\delta$ is a fluctuation term around zero, we can rewrite it in a transparent form as

\begin{equation}
\delta_{ij}(t)=2\Delta_{ij}\cos\left(t\sqrt{\frac{\tau}{\tau^{\prime}}}+\phi\right)
\end{equation}
where $\Delta_{ij}\sim \Delta$ represents the fluctuation amplitude. Replacing this form directly into \eqref{integ} and integrating over $\phi\in [0,2\pi]$, the oscillating terms disappear, and we
achieve an estimation for the amplitude $\Delta$, which is
\begin{equation}
2\Delta^{2}\frac{1+\tau^{\prime}/\tau}{\tau^{\prime}/\tau}=\int_{0}^{\tau^{\prime}}\left(-\frac{\tau}{\tau^{\prime}}J_{ij}^{(Hebb)}+\frac{\tau}{\tau^{\prime}}\xi_{i}(t)\xi_{j}(t)\right)^{2}dt.
\end{equation}
Expanding the square in the r.h.s. and assuming $\tau'/\tau\gg1$, we can write down the following estimation:
\begin{equation}
\begin{split} &
	\label{eq:fluct} \Delta \approx\sqrt{\frac{1}{2}\frac{\tau}{\tau^{\prime}}\left[1-\left(J_{ij}^{(Hebb)}\right)^{2}\right]}.
\end{split}
\end{equation}
Thus the fluctuation amplitude around the Hebbian
kernel is proportional to ${(\tau/\tau')}^{1/2}$. In Fig. \ref{converg} we plotted in dashed lines this estimation for the fluctuations around the Hebbian kernel, and we find a perfect agreement between the theoretical prediction \eqref{eq:fluct} and the numerical simulations. For general $\beta$, the same analysis allows us to quantify that
\begin{equation}\label{eq:beta_fluct}
	\Delta  (\beta )\simeq \sqrt{\Delta ^2 \tanh ^2 \beta + [J_{ij}^{(Hebb)}(1-\tanh \beta )]^2},
\end{equation}
with $\Delta $ being the $\beta\to\infty$ evaluation given by \eqref{eq:fluct}. The comparison between this prediction and the numerical analysis can be found in Fig. \ref{converg} (panel d) where the theoretical prediction is represented by the dashed line.

\section*{Acknowledgments}
This work is supported by Ministero degli Affari Esteri e della Cooperazione Internazionale (MAECI)  via the BULBUL grant (Italy-Israel), CUP Project n. F85F21006230001.
\newline
EA and AF acknowledge Progetto Ricerca Ateneo (RM120172B8066CB0).

\bibliographystyle{ieeetr}
\bibliography{AABFM_V4}

\end{document}